\documentclass[11pt,twoside]{article}
\usepackage{asp2010}

\resetcounters

\begin{document}

\title{Exploring the effects of detailed chemical profiles on
  the\\ adiabatic oscillation spectrum of sdB stars: First Results}
\author{M. M. Miller Bertolami,$^{1,2}$ A. H. C\'orsico,$^{1,2}$ and
  L. G. Althaus$^{1,2}$ \affil{$^1$ Facultad de Ciencias
    Astron\'omicas y Geof\'isicas,\\ Universidad Nacional de La Plata,\\
    Paseo del Bosque s/n, B1900FWA, La Plata, Argentina} \affil{$^2$
    CONICET, Argentina}}

\begin{abstract}
We present results of an ongoing study of the pulsational properties
of sdB stellar models aimed at exploring the consequences of detailed
chemical transitions for radial, $p-$ and $g-$ modes.  In particular,
we focus on the effects of diffusion at the H-He transition and of
He-burning at the convective cores. 

We find that diffusion of He and H has a strong impact on the period
spectrum of sdBVs stars, leading to less efficient mode trapping. Our
results also suggests that asteroseismology of sdBVs stars might offer
a very good opportunity to constrain extramixing processes in the
He-burning cores of horizontal branch stars.

\end{abstract}

\section{Introduction}
Hot subdwarf stars (sdO, sdB) configure an ubiquitous population of
stars, which are located in the HR-diagram between the main sequence
and white dwarf stars. Most hot subdwarfs are supposed to be He-core
burning stars with H-envelopes which are too thin to sustain
H-burning. Thus, hot subdwarf stars are identified with stars in the
extreme horizontal branch (EHB, see \citealt{2009ARA&A..47..211H} for
an excellent review). While the fate of hot subdwarfs is very clear,
they will evolve towards the white dwarf stage avoiding the asymptotic
giant branch, their origin is not fully understood. Most hot subdwarf
stars are supposed to be formed from red giants that lost almost the
entire envelope (either due to close binary interaction, ingestion of
substellar companions or rotationally enhanced winds) or He white
dwarf mergers \citep{2002MNRAS.336..449H,2003MNRAS.341..669H}.

\begin{figure}[ht]
\begin{center}
\includegraphics[clip, angle=0, width=12cm]{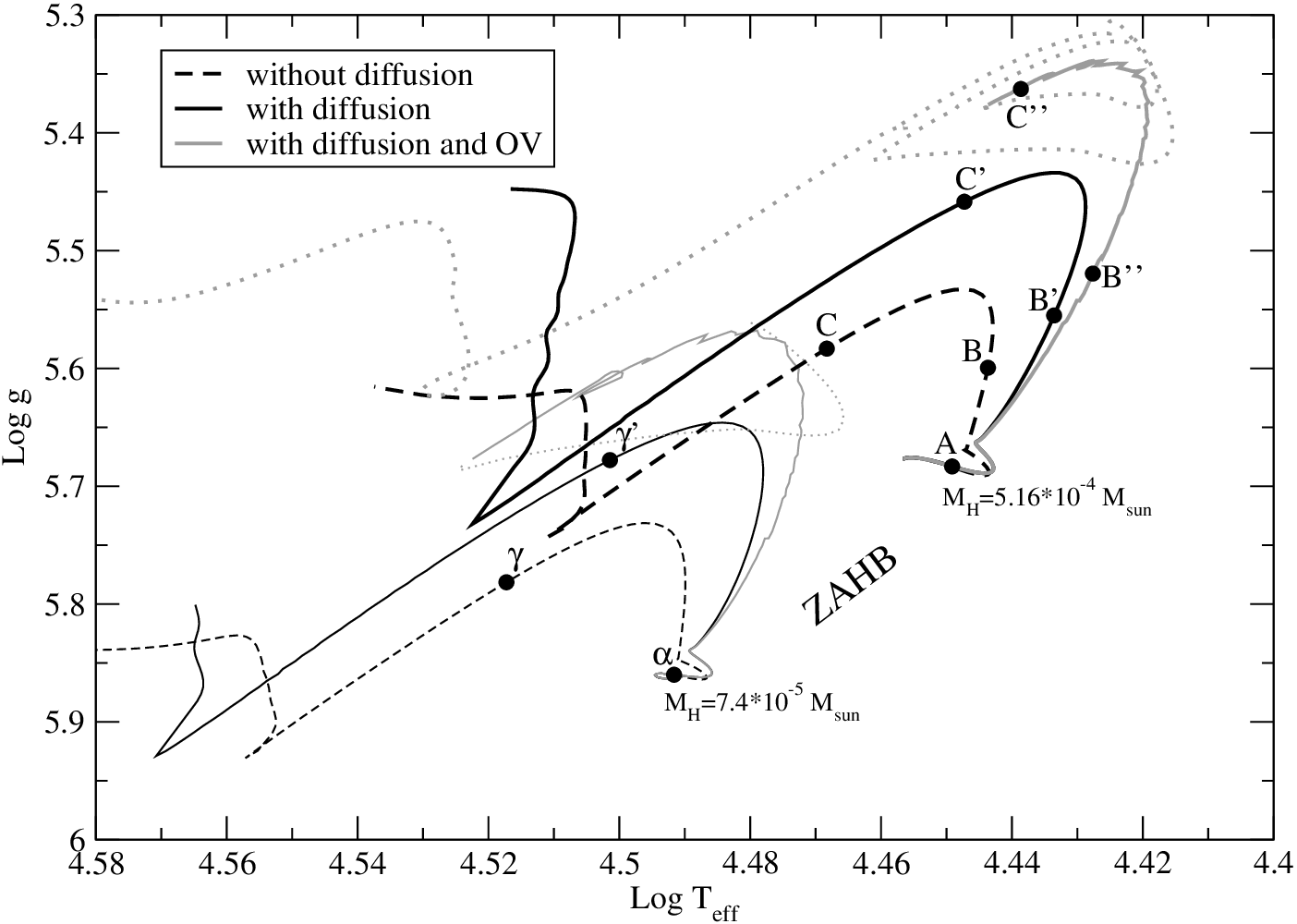}
\caption{Location in the $T_{\rm eff}-g$ diagram of the six sequences
  discussed in this work.  Black dots indicate particular models
  discussed through the text. In particular models A and $\alpha$
  correspond to initial ZAHB models for the two different H-envelope
  masses adopted (thick; $M_{\rm H}=5.16\times 10^{-4}M_\odot$, thin;
  $M_{\rm H}=7.40\times 10^{-5}M_\odot$).}
\label{Fig:HR} 
\end{center}
\end{figure}

The discovery of multiperiodic pulsations in some sdB stars opened the
oportunity to sound the interior of hot subdwarf stars with
asteroseismological tools.  Specifically two main families of
pulsators have been discovered within the sdB stars, the rapid
pulsators (sdBVr; \citealt{2010IBVS.5927....1K}) discovered by
\cite{1997MNRAS.285..640K}, and the slow pulsators (sdBVs;
\citealt{2010IBVS.5927....1K}) discovered by
\cite{2003ApJ...583L..31G}. While sdBVr stars show short pulsation
periods ($\sim 80-400$ s) ascribed to radial modes and non-radial
$p$-modes, sdBVs pulsations (with periods $\sim 2500-7000$ s) are
associated to non-radial long period $g$-modes.  Pulsations in both
groups of variable stars have been explained by the action of the
$\kappa$-mechanism due to the partial ionization of iron group
elements in the outer layers, where these elements are enhanced by the
action of radiative levitation \citep{1997ApJ...483L.123C,
  2003ApJ...597..518F}.  Besides these two main groups, two other types of
pulsating stars have been found among hot subdwarfs, the sdOV stars
\citep{2008A&A...486L..39F, 2011ApJ...737L..27R} with pulsations also
driven by the $\kappa$-mechanism and the only He-rich sdBV star (LS
IV-14$^\circ$116, \citealt{2005A&A...437L..51A}) which has been
recently proposed to be the first observed star with pulsations driven
by the $\epsilon$-mechanism \citep{2011ApJ...741L...3M}.

In the present paper we communicate first results of an ongoing study
aimed at understanding the effects of detailed chemical structures on
the adiabatic period spectrum of sdB stars.

\section{Stellar evolution models and numerical details}

The sequences of stellar models presented in this work were computed
with {\tt LPCODE}, a numerical code for solving the equations of
stellar evolution, which was already used to model the formation of
He-rich subdwarf stars within the hot-flasher scenario
\citep{2008A&A...491..253M}. {\tt LPCODE} is a Henyey-type stellar
evolution code designed specifically to compute the whole evolution of
low and intermediate mass stars and is described extensively in
\cite{2005A&A...435..631A} and references therein. Therefore, in
what follows, we only refer to the code to mention some particular
features of special interest to this work.

 In the present work, initial Zero Age Horizontal Branch (ZAHB) models
 are full evolutionary structures which have been evolved through the
 helium core flashes at the end of the Red Giant Branch (RGB)
 \citep{2008A&A...491..253M}. Consequently the models presented in
 here are only representative of canonical post-He-flash sdB stars
 which have a significantly different interior structure to
 post-non-degenerate sdB models (see \citealt{2008A&A...490..243H}).
 In particular we have followed the evolution of initially $Z=0.02$
 and $M=0. 47426 M_{\odot}$ stellar models with two different
 H-envelope masses ($M_{\rm H}$; see Fig. \ref{Fig:HR}) through the
 He-core burning stage. We computed sequences of models from the ZAHB
 to the Terminal Age Horizontal Branch (TAHB) under three different
 physical assumptions (see Fig.  \ref{Fig:HR}): 1- Standard stellar
 evolution models used for reference (no element diffusion, no core
 overshooting), 2- More realistic stellar evolution models which
 include the effects of element diffusion, 3- Including the possible
 consequences of core overshooting at the convective core (also
 including the effects of element diffusion).  Element diffusion was
 considered only for H, $^3$He, $^4$He, $^{12}$C, $^{13}$C, $^{14}$N
 and $^{16}$O and computed under the assumption of complete ionization
 and considering the effects of gravitational settling, thermal
 diffusion and chemical diffusion but neglecting the effects of
 radiative levitation (for similar computations that include the
 effects of radiative levitation of Fe and Ni see
 \citealt{2011MNRAS.tmp.1435H}). Our treatment of time-dependent
 element diffusion is based on the multicomponent gas picture of
 \cite{1969fecg.book.....B}. Specifically, we solved the diffusion
 equations within the numerical schemes described in
 \cite{2000MNRAS.317..952A} and \cite{2005A&A...435..631A}.
 Overshooting was treated as an exponentially decaying diffusive
 process with a free parameter $f=0.015$, following
 \cite{1997A&A...324L..81H}.

Radial and non radial adiabatic oscillations were computed with an
updated version of the pulsation code described in
\cite{2006A&A...454..863C}.

\section{$g$-modes: Effects of H/He-diffusion and core overshooting.}
\begin{figure}
\begin{center}
\includegraphics[clip, angle=0, width=12cm]{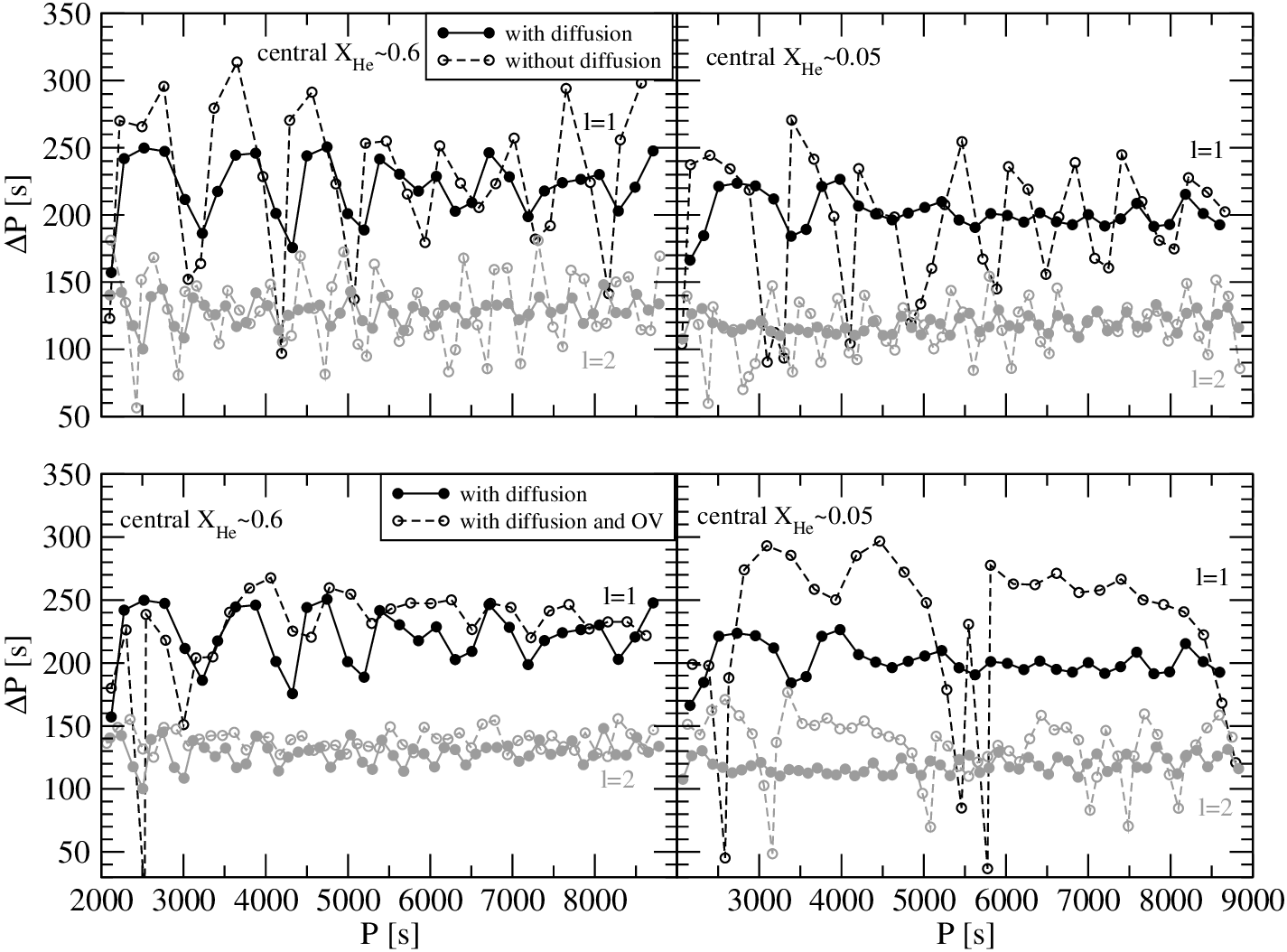}
\caption{{\it Upper Panels:} Period spacing properties of $g$-modes
  for the models with (solid lines) and without (dashed lines)
  diffusion for $\ell=1$ (black) and $\ell=2$ (grey). Left panel
  corresponds to models B and B' early on the HB, when the central
  content of He is still high ($X_{\rm He}\sim 0.6$), while right
  panel corresponds to models C and C', close to the TAHB ($X_{\rm
    He}\sim 0.05$). Note the nearly equally spaced periods that occur
  in models that include diffusion due to the absence of strong mode
  trapping at the H-He transition. {\it Lower Panels:} Same as upper
  panels but comparing models with and without overshooting (OV) at
  the He-burning core (right panel: models B' and B'', left panel:
  models C' and C''). Note the strong deviations from uniform spacing
  that occur in the model that includes overshooting due to the
  complex chemical profiles at the outer boundary of the He-core
  (lower right panel). }
\label{Fig:dP-P} 
\end{center}
\end{figure}

\begin{figure}
\begin{center}
\includegraphics[clip, angle=0, width=10cm]{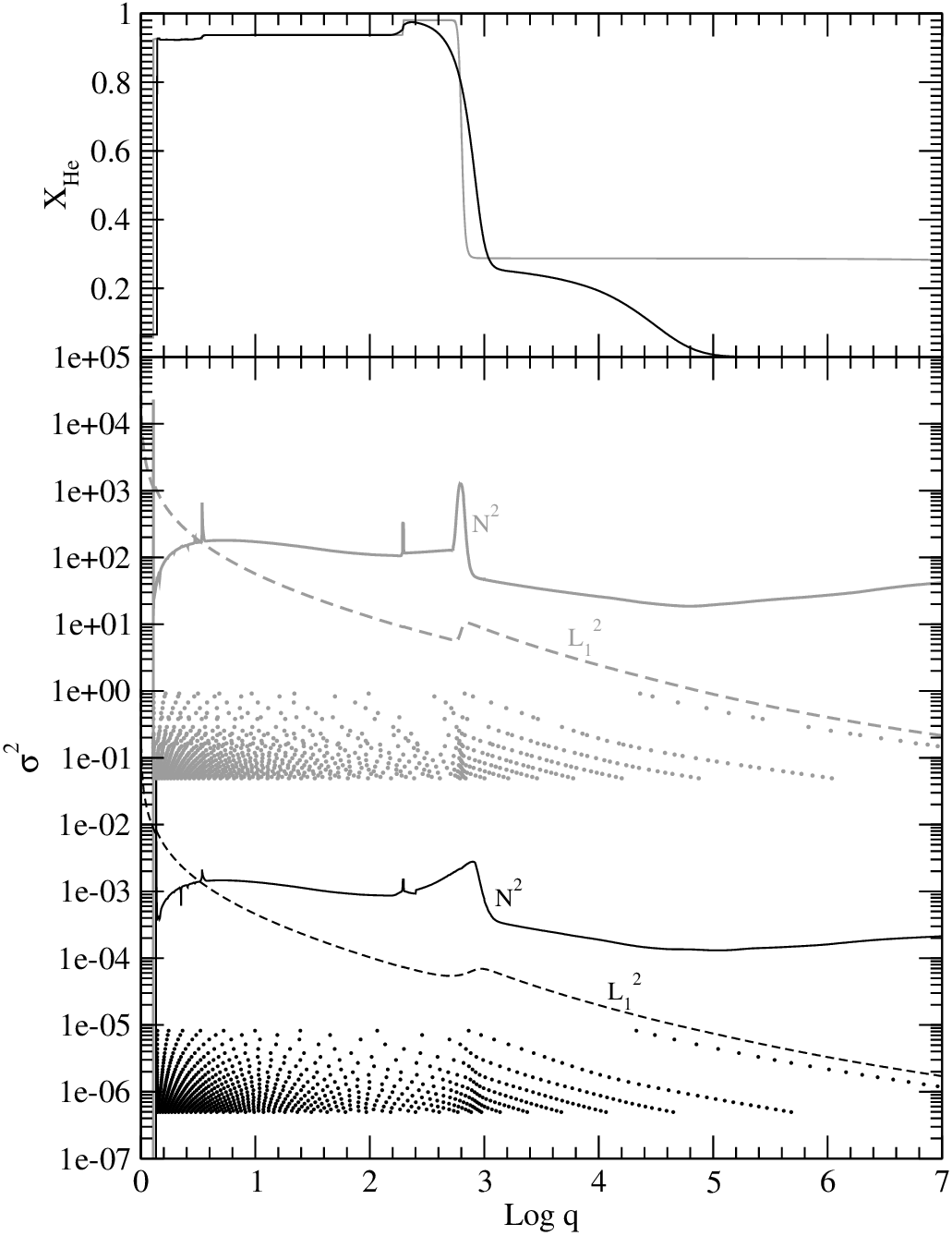}
\caption{Properties of the models with (black) and without (grey)
  diffusion close to the TAHB (models C, C'). {\it Upper Panel:} He profiles
  of the models. Note the steep profile present in the model without
  diffusion which causes strong mode trapping. {\it Lower Panel:}
  Brunt-V\"ais\"al\"a and Lamb frequencies of the models. Dots
  indicate the location of the nodes of adiabatic modes at different
  frequencies within the range of periods shown in
  Fig. \ref{Fig:dP-P}. Frequencies for the model without diffusion
  have been arbitrarily multiplied by $10^5$ to avoid overlapping. Note
  the strong mode trapping that is apparent at the H-He transition in
  the model without diffusion (grey).}
\label{Fig:BV-diff} 
\end{center}
\end{figure}

The upper panels of Fig. \ref{Fig:dP-P} display the period spacing
against period properties of models computed from a same initial model
(model A, see Fig. \ref{Fig:HR}) but under different assumptions
during the He-core burning evolution.  Specifically, we considered two
sequences, one that includes the effects of H/He-diffusion in the
envelope and one without any diffusion. In agreement with
\cite{2009A&A...508..869H} H/He-diffusion leads to a broadening of
composition gradients and less efficient mode trapping (see
Fig. \ref{Fig:BV-diff}), as found to happen in pulsating white dwarfs
\citep{2001A&A...380L..17C}. The impact of H/He-diffusion on the
Brunt-V\"ais\"al\"a frequency can be clearly appreciated in the lower
panel of Fig. \ref{Fig:BV-diff} in the broadening and flatening of the
outermost peak, which is located at the base of the H-rich
envelope. Notice also the effect of the H-He transition in the
location of the nodes of the mode eigenfunctions
(Fig. \ref{Fig:BV-diff}). Note that the effect of diffusion is already
apparent early on the horizontal branch evolution, by the time only
one third of the original He has been burnt ---He$_{\rm central}\sim
0.6$, Fig.  \ref{Fig:dP-P}, upper left panel, models B and B' of Fig.
\ref{Fig:HR}.  The effect of diffusion is even more noticeable close
to the TAHB, by the time the central He content has dropped to
He$_{\rm central}\sim 0.05$ ---Fig.  \ref{Fig:dP-P}, upper right
panel, models C and C' of Fig. \ref{Fig:HR}. While the model evolved
with diffusion displays almost no trapping features, the effects of
mode trapping in the $\Delta P-P$ values of the model evolved without
diffusion are apparent. It is clear that the effects of H and He
diffusion on the chemical profile of sdB stars can not be neglected
for detailed asteroseismological fits of observed periods in sdBVs
stars. In this connection, it is worth noting the recent determination
by \cite{2011MNRAS.414.2885R} that mode trapping in real stars must be
substantially lower than what static models without H/He-diffusion
\citep{2010ApJ...718L..97V} indicate.  Then the results presented by
\cite{2011MNRAS.414.2885R} might be just indicating the need to
include H/He-diffusion in the structures used in asteroseismological
determinations of sdB stars (see also Hu et al., Reed et al. these
proceedings). If this is so, fully evolutionary models which include
diffusion in a selfconsistent way (like those already presented by
\citealt{2011MNRAS.tmp.1435H}) might be necessary in order to perform
asteroseismological studies of sdBV stars.

\begin{figure}
\begin{center}
\includegraphics[clip, angle=0, width=10cm]{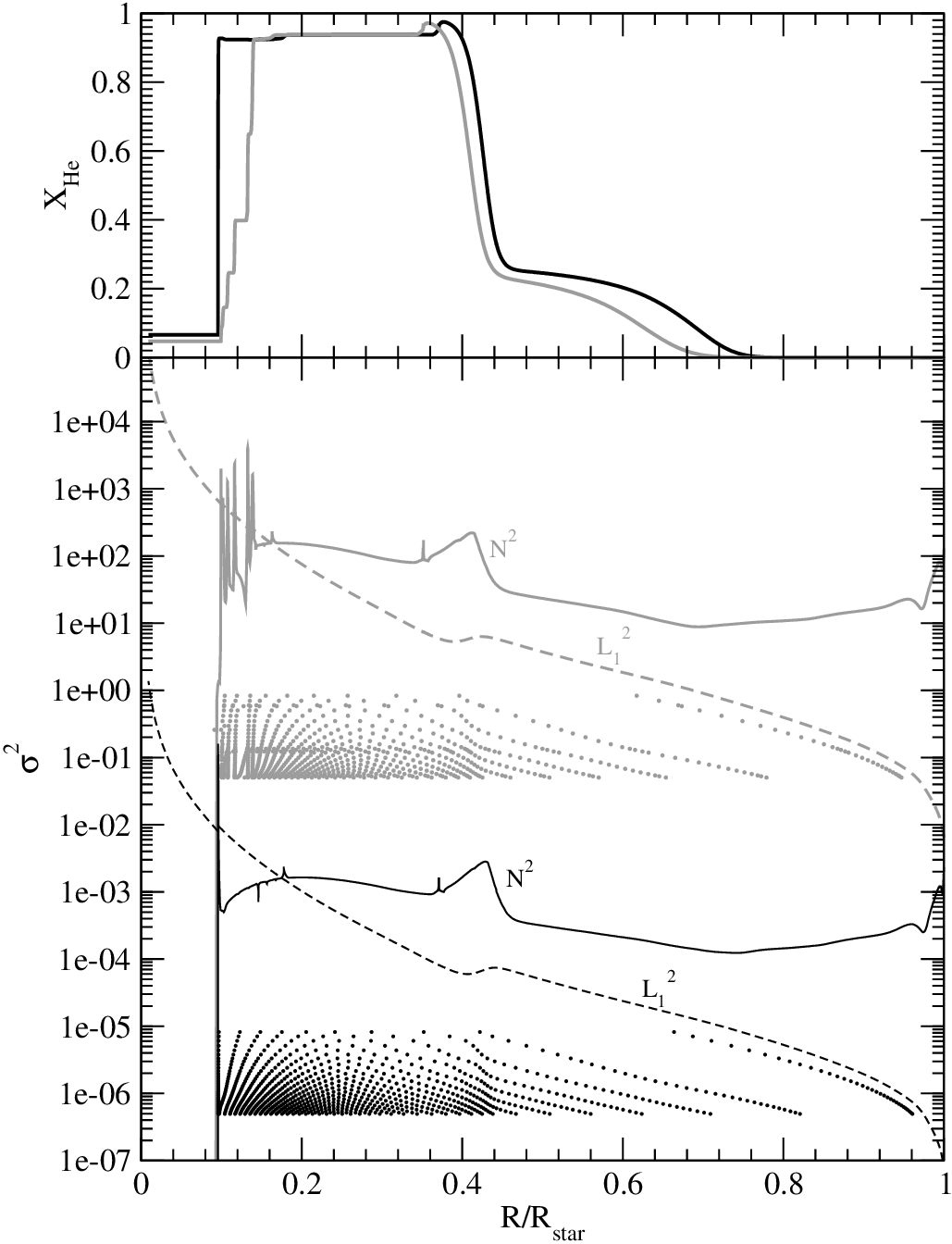}
\caption{Same as Fig. \ref{Fig:BV-diff} but for the model (with
  diffusion) that includes exponentially decaying overshooting
  compared with the model that only includes diffusion but no
  extramixing at the convective core (models C'and C''). Note the
  complex He chemical profile at the outer boundary of the He-burning
  core (grey, upper panel) that leads to a very complex structure of
  the Brunt-V\"ais\"al\"a frequency (lower panel). Note the mode
  trapping caused by these chemical transitions which can be
  appreciated in the location of the nodes of the model with
  overshooting (grey dots).}
\label{Fig:BV-OV} 
\end{center}
\end{figure}

In order to explore to what extent the core chemical structure and
size impact mode trapping and, thus the period spacing of sdBVs stars,
we computed models which include, in addition to the effects of
diffusion, overshooting at the convective core. Besides staying longer
in the horizontal branch and being more luminous due to a larger
He-burning convective core, models that include exponentially decaying
overshooting also develop complex step-like chemical profiles at the
outer convective boundary (see upper panel of
Fig. \ref{Fig:BV-OV}). These chemical transitions have a strong impact
on the Brunt-V\"ais\"al\"a frequency ($N$, e.g.  Fig. \ref{Fig:BV-OV}
lower panel). Note, in particular, that $g$-modes in the range of
periods observed in sdBVs stars have several nodes in the region of
the star where this step-like chemical profile is located
(Fig. \ref{Fig:BV-OV} lower panel). Thus, we expect the $g$-mode
period spacings of our model to be sensitive to the details of the
chemical transition at the outer boundary of the CO-rich core.  This
is clearly shown by the lower panels of Fig. \ref{Fig:dP-P} where we
compare the $\Delta P-P$ values of the models evolved with
overshooting (B'', C'' which have a complex Brunt-V\"ais\"al\"a
frequency at the core boundary) and without overshooting (B', C' which
has a relatively simple Brunt-V\"ais\"al\"a frequency at the core
boundary). Then, it might be possible to use periods of sdBVs stars to
constrain not only the location (as already done by
\citep{2010ApJ...718L..97V}) but also the shape of the outer boundary
of the He-burning convective core. It is then worth noting that, as
the extent of extramixing processes at the He-burning convective core
of horizontal branch (HB) stars is not well constrained, sdBVs
asteroseismology might be extremely useful tool to learn about the
convective cores of the whole population of HB stars.

\section{$p$-modes and radial modes: H-He diffusion and realistic ZAHB H
profiles.} 
\begin{figure}
\begin{center}
\includegraphics[clip, angle=0, width=12cm]{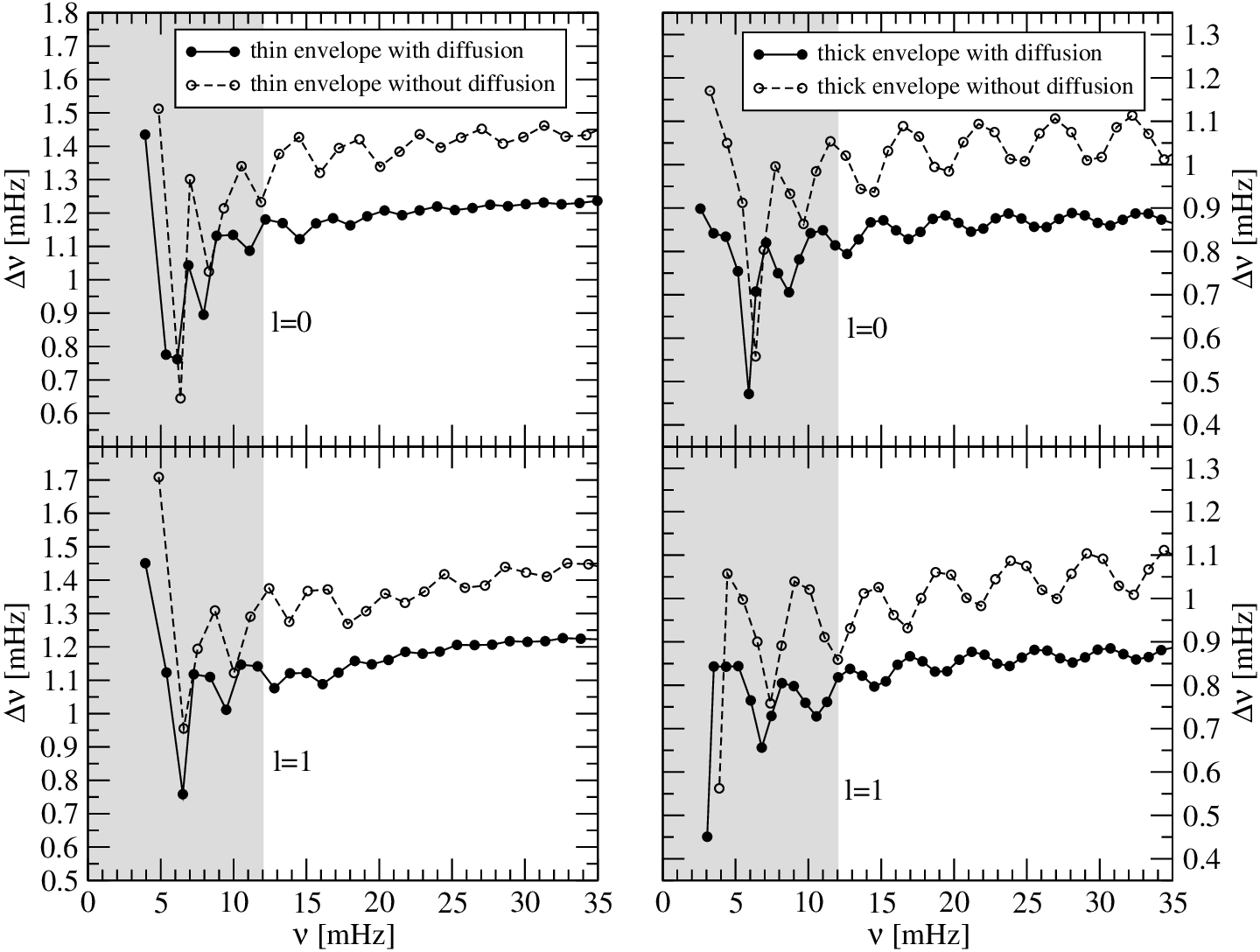}
\caption{Frequency spacing properties for the models with (solid
  lines) and without diffusion (dashed lines) close to the TAHB
  (models $\gamma$, $\gamma$', C and C'). Upper (lower) panels
  correspond to radial modes ($l=1$ $p$-modes). Values for the models
  with thick envelopes (C, C') are shown in the right panels while
  the left panels display the values for the models with thin H-envelopes
  ($\gamma$, $\gamma$'). The grey area correspond to the frequencies
  observed in sdBVr pulsators.}
\label{Fig:df-f} 
\end{center}
\end{figure}

 Fig. \ref{Fig:df-f}  diplays the $\Delta\nu$-$\nu$
properties for $\ell=0$ and $\ell=1$ modes of models close to the end
of the TAHB (Fig.  \ref{Fig:HR} , models C, C', $\gamma$, $\gamma$')
computed with and without chemical diffusion and with two different
H-envelope thicknesses. As can be clearly appreciated, in both panels
of Fig \ref{Fig:df-f} , microtrapping features are significantly
eroded at high frequencies and the effect is even noticeable for low
order acoustic modes within the range of periods of sdBVr pulsators
($\nu < 12$ mHz). This is caused by the smoothing of the H-He
transition which affects not only the Brunt-V\"ais\"al\"a frequency
but also Lamb frequencies ($L_{\ell}$) and thus both $p$ and radial
modes.

\begin{figure}
\begin{center}
\includegraphics[clip, angle=0, width=10cm]{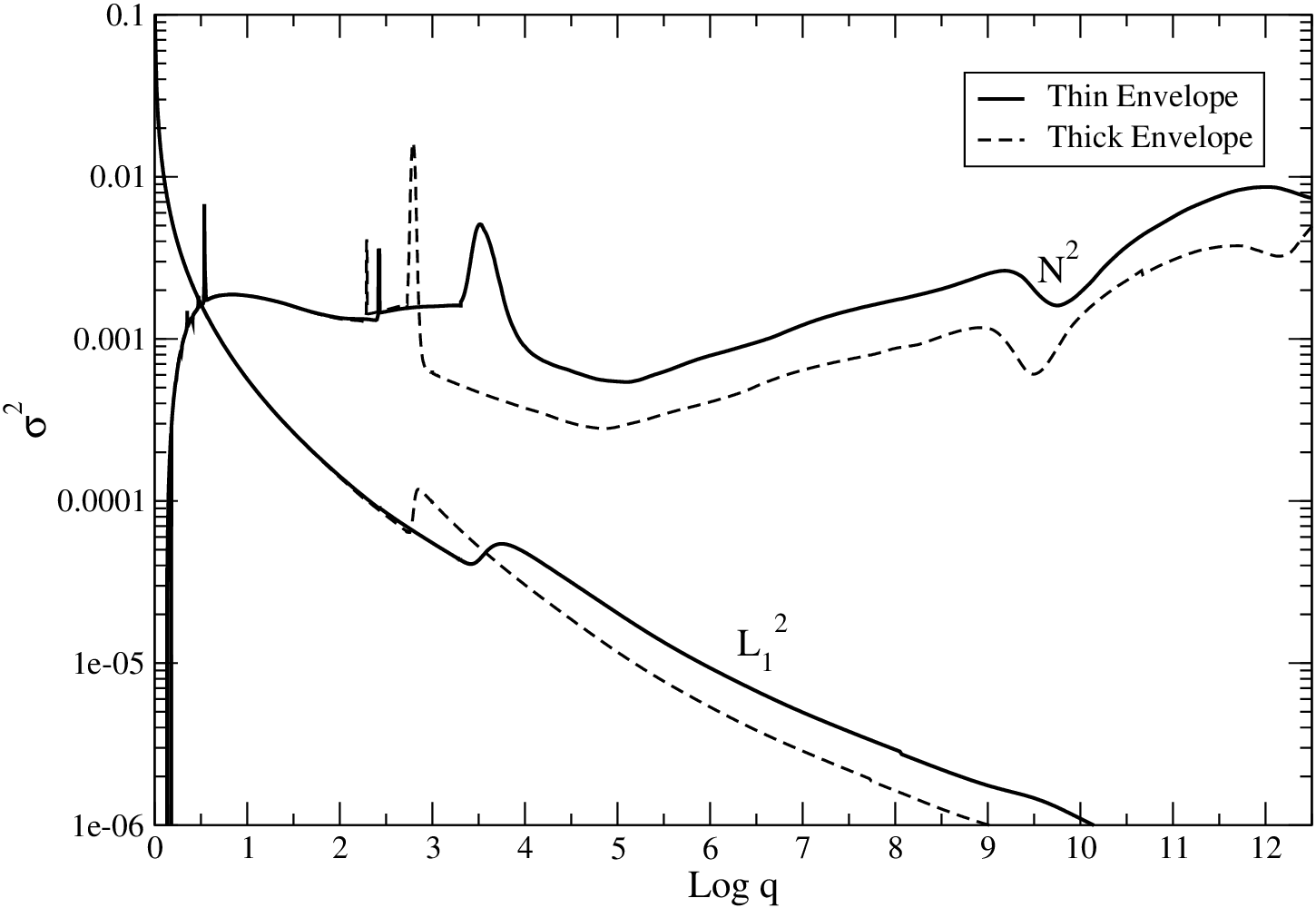}
\caption{ Brunt-V\"ais\"al\"a and Lamb frequencies of initial ZAHB
  models with thin and thick H-envelopes (A and $\alpha$). Note the
  higher peak in the Brunt-V\"ais\"al\"a frequency of the model with a
  thick H-envelope which will cause stronger mode trapping in thick
  envelope models even in the absence of diffusion.}
\label{Fig:BV-del-gru} 
\end{center}
\end{figure}
In connection with the trapping features of acoustic modes it is worth
noting that stellar evolution computations predict that the shape of
the H-He transition depends on the thickness of the H-rich envelope
already at the ZAHB (see Fig. \ref{Fig:BV-del-gru} , models A and $\alpha$ of
Fig. \ref{Fig:HR}). Thus, contrary to usual assumptions (e.g. Fig. 33
of \cite{2002ApJS..139..487C}), thin envelope models not only have a
more external peak in the Brunt-V\"ais\"al\"a frequency but also a
smoother peak than thick envelope models. This is true despite the
fact that the shape of the chemical transitions are equal in both
models in the Lagrangian coordinate ($m_{(r)}$) and is caused by the more
expanded envelope of the models with thinner envelopes.

The impact of the effect of detailed chemical H-He transitions on
seismic envelope mass determinations of sdBVr stars should be
explored.

\section{Conclusion}

We have explored the effects of detailed chemical transitions arising
from evolutionary processes on the adiabatic oscillation spectrum of
sdBV stars. 

Our results show that long period, $g$-mode, oscillations
as those observed in sdBVs stars are affected by H/He-diffusion. In
particular we find that the models that include diffusion have periods
which are more equally spaced than models which do not include
diffusion at the H-He transition. In this context the recent finding
of nearly equally spaced periods in sdBVs stars
\citep{2011MNRAS.414.2885R} might be a strong indication of the
occurrence of diffusion at the H-He transition of sdB stars. 

Our results also indicate that the details of the chemical transitions
at the outer boundary of the He-burning core may affect the mode
trapping features in sdBVs stars. This might open a very good
oportunity to constrain both the size of the core and the shape of the
chemical profiles at the outer boundary of the He-burning core.  As
extramixing processes at the boundaries of He-burning cores are not
well understood or constrained, asteroseismology of sdBVs stars might
offer the opportunity to learn about extramixing processes in the
cores of horizontal branch stars. 

Finally, our preliminary study suggests that the details of the H-He
transition might also affect the frequencies of modes in the observed
frequency range of sdBVr stars.  We think that the impact of realistic
H-He transitions in asteroseismological determinations of sdBVr stars
should be explored.

\acknowledgements This work has been supported by grants PIP
112-200801-00904 and PICT-2010-0861 from CONICET and ANCyT,
respectively. M3B thanks both the Varsavsky Foundation and the
organizers of the Fifth Meeting on Hot Subdwarf Stars \& Related
Objects for the financial assistance that allowed him to attend the
meeting.

\end{document}